\documentclass[11pt,a4paper]{JHEP3}
\usepackage{amsmath, amssymb}
\usepackage{euscript}
\def\be{\begin{eqnarray}}
\def\ee{\end{eqnarray}}
\def\0{\nonumber}
\newcommand\EE{\EuScript{E}}
\newcommand\EU{\EuScript{U}}
\usepackage{amsfonts}
\usepackage{verbatim}
\newcommand\N{{\cal N}}
\newcommand\U{{\cal U}}
\newcommand\E{{\cal E}}

\def\k{\kappa}

\def\m{\bar m}
\def\n{\bar n}

\preprint{SISSA/101/2007/EP\\\tt hep-th/yymm.nnnn}

\title{Spectral properties of ghost Neumann matrices}

\author{ L.Bonora\\
International School for Advanced Studies (SISSA/ISAS)\\
Via Beirut 2--4, 34014 Trieste, Italy, and INFN, Sezione di
Trieste\\
E-mail:   \email{bonora@sissa.it},}

\author{R.J.Scherer Santos\\
Centro Brasileiro de Pesquisas Fisicas (CBPF-MCT)-LAFEX\\
R. Dr. Xavier Sigaud, 150 - Urca - Rio de Janeiro - Brasil - 22290-180\\
E-mail: \email{scherer@cbpf.br},}

\author{D.D.Tolla\\
Center for Quantum SpaceTime (CQUEST), Sogang University\\
Shinsu-dong 1, Mapo-gu, Seoul, Korea\\
E-mail:  \email{tolla@sogang.ac.kr}}

\abstract{We continue the analysis of the ghost wedge states in 
the oscillator formalism by studying the spectral properties of
the ghost matrices of Neumann coefficients. We show that the traditional spectral
representation is not valid for these matrices and propose a new
heuristic formula that allows one to reconstruct them from the knowledge of
their eigenvalues and eigenvectors. It turns out that additional data, which
we call boundary data, are needed in order to actually implement the
reconstruction. In particular our result lends support to the 
conjecture that there exists a ghost three strings vertex with properties 
parallel to those of the matter three strings vertex.}

\keywords{String Field Theory, Ghost Wedge States, Star Product}

\begin{document}

\maketitle

\section{Introduction}

This paper is complementary to the analysis, started in \cite{BMST}, 
of the conjectured equivalence 
\be
e^{ - \frac {n-2}2\left({\cal L}^{(g)}_0 +
{\cal L}_0^{(g)\dagger}\right)}|0\rangle
= {\cal N}_n\, e^{ c^\dagger S_n b^\dagger}|0\rangle\equiv |n\rangle 
\label{ghostwedge}
\ee
which is a crucial one in the recent developments in open string field theory
\cite{Schnabl05,Okawa1,Ellwood:2006ba,RZ06,ORZ,Schnabl:2007az,
KORZ,Fuchs2,Fuchs3,Fuchs0,Fuchs1,Okawa2,Okawa3}.
Here $|n\rangle$ are the ghost wedge states in the oscillator formalism
\cite{Samu,CST,GJ1,GJ2,Ohta} , 
which of course must coincide with the corresponding surface wedge states.
In \cite{BMST} we dealt with the LHS of this equation. We showed that, if
we understand it ordered according to the natural normal ordering, it
can be cast into the midterm form 
in (\ref{ghostwedge}), and we diagonalized the matrix $S_n$ in such a
squeezed state. Then we proved that, {\it if} we are allowed to 
star--multiply the squeezed states representing the ghost wedge states 
$|n\rangle$  the same way we do for the matter wedge 
states and diagonalize the corresponding matrices, the eigenvalue we obtain 
in the two cases are the same. In this paper we focus on the spectral properties
of such operators like $S_n$ and, among other things, 
we exhibit evidence that the above {\it if} is justified.

To be more precise, in order to fully prove (\ref{ghostwedge}) we have 
two possibilities. The most direct alternative is to
define the three strings vertex for the 
ghost part, and thus the star product, pertinent to the natural normal  
ordering in the oscillator formalism; then to construct the wedge states
appropriate for this vertex; finally to diagonalize the latter and 
shows that they coincide with the midterm of (\ref{ghostwedge}) (with some 
additional specifications that will be clarified in due course).
Unfortunately the construction of the ghost vertex is not so straightforward
as one would hope. Relying on the common lore on this subject, we 
face a large number of possibilities, which are mostly linked to the
ghost zero mode insertions and our attempts in this direction so 
far have been unfruitful. Before continuing in such a challenging program
it is wise to gather some evidence that the vertex one is looking for 
does exist and some indirect information about it. This is
the original motivation of the present paper, which  
relies on the second alternative.
 
Having diagonalized the matrices $S_n$ in the midterm of (\ref{ghostwedge})
in the basis of weight 2 differentials, see \cite{BMST}, one may wonder 
whether one can reconstruct the original matrices. For the matter part this 
is a standard procedure, 
simply one uses the spectral representation of the infinite
matrices involved. But for the ghost sector we are interested in here things
are more complicated (it should be recalled that the infinite matrices 
$S_n$ are not square
but {\it lame}, i.e. infinite rectangular). 
Ultimately, the answer is: 
yes, we can reconstruct the $S_n$ matrices; in other words, we can derive
the RHS of (\ref{ghostwedge}) from the LHS, but the procedure is more 
involved than in the matter case. In fact the traditional spectral
representation is not valid for lame matrices and we have to figure out a new
heuristic formula that allows us to reconstruct them from  
their eigenvalues and eigenvectors. It turns out that additional data, which
we call boundary data, are needed in order to actually implement the
reconstruction. Once this is done we can extract
from them basic information about the Neumann coefficients
matrices of the ghost three strings vertex. 

The main results of our paper are the study of the spectral properties
of the infinite matrices $S_n$ in the $b-c$ ghost bases, the reconstruction
recipe for such infinite matrices (which is an interesting result in itself) 
and the evidence for the existence of
the three--strings vertex we need for the ghost sector in the 
natural normal ordering.

The paper is organized as follows. In section 2, which is essentially 
pedagogical, we present an example of three strings vertex which is not
the one we are looking for as it cannot be diagonalized in the weight 
2 basis, but has all the other good properties we expect of the true vertex.
This example also illustrates the problems one comes across in constructing
a ghost three strings vertex. In section 3 we make contact with the results
of \cite{BMST} and give a more detailed proof that the squeezed states 
in the midterm of (\ref{ghostwedge}) have the same eigenvalue as the 
ghost wedge states in the oscillator formalism. We clarify
that this is not enough to prove that (\ref{ghostwedge}) holds, and,
in section 4, we show where the problem lies and propose a new heuristic
formula for the reconstruction of infinite lame matrices. Finally section 5
is devoted to our conclusions. Three Appendices contain details of the
calculations needed in the text. In particular Appendix C presents a new proof
of the fundamental eq.(\ref{eigenA=0}).

{\bf Notation}. Any infinite matrix we meet in this paper is either square short 
or long legged, or lame. In this regard we will often use a compact 
notation: a subscript $_s$ will represent an integer label $n$ running 
from $2$ to $\infty$, while a subscript $_l$ will represent a label 
running from $-1$ to $+\infty$. So $Y_{ss}, Y_{ll}$ will denote
square short and long legged, respectively; $Y_{sl}, Y_{ls}$ will 
denote short--long and long--short lame matrices, respectively.
With the same meaning we will say that a matrix is $(ll),(ss),(sl)$ or 
$(ls)$.
In a similar way we will denote by $V_s$ and $V_l$ a short and long 
infinite vector, to which the above matrices naturally apply.
Moreover, while $n,m$ represent generic matrix indices, at times we will
use $\n,\m$ to stress that they are short, i.e. $\n,\m\geq 2$.

\section{The three strings vertex}

This section is mostly pedagogical. We would like here to explain
what are the problems with defining a three strings vertex for the
ghost sector that fits the purposes of proving eq.(\ref{ghostwedge})  
The first problem we have to face is normal ordering. 
We will have in mind two main cases of normal orderings,
those we have called {\it natural} and {\it conventional normal ordering} 
in \cite{BMST}. The former is the obvious normal ordering required when 
the vacuum is $|0\rangle$, the latter is instead appropriate to the vacuum 
state $c_1|0\rangle$. A consistent vertex for conventional normal ordering 
exists, is the one explicitly computed by Gross and Jevicki, \cite{GJ2},
who used the vacuum $c_0c_1|0\rangle$ (for general problems connected with
the ghost sector, see \cite{GRSZ1,HKw,Oku2,Kishimoto}).
But it is not  
what we need in the natural normal ordering case. A second problem is 
generated by the ghost insertions, 
which are free and there is no a priori principle to fix them.
We know however that a certain number of conditions should be satisfied. 
One is BRST invariance of the three strings vertex. This is unfortunately
hard to translate into a practical recipe for construction. Other
conditions, i.e. cyclicity, $bpz$-- compatibility and commutativity of 
the Neumann coefficients matrices are more useful from a constructive 
point of view.
 
In the sequel we will consider a definite example. Even though it turns out
not to be the right vertex we are looking for, it will allow us to
illustrate many questions which would sound rather abstruse in the abstract. 
 
To start with we first recall the relevant anti--commutator and $bpz$ rules
\be
[c_n,b_m]_+=\delta_{n+m,0}, \quad \quad bpz(c_n) = -(-1)^n c_{-n},
\quad\quad bpz(b_n) = (-1)^n b_{-n}
\0\ee
The we define the state $|\hat 0\rangle = c_{-1}c_0c_1|0\rangle$, 
where $|0\rangle$ is the SL(2,R)--invariant vacuum, the
tensor product of states
\be
_{123}\langle \hat\omega|= {}_1\langle \hat 0|_2\langle \hat 0 |_3\langle 0|
\label{omegahat}
\ee
carrying total ghost number 6, and
\be
|\omega\rangle_{123} &=& | 0\rangle_1| 0\rangle_2|\hat 0\rangle_3
\label{omega}
\ee
carrying total ghost number 3. They satisfy 
$_{123}\langle \hat\omega|\omega\rangle_{123}=1$. Finally
we write down the general form of the three strings vertex  
\be
\langle \hat V_{3}|=\,{\cal K}\,{}_{123}\langle \hat \omega|
e^{\hat E}, \quad\quad
\hat E=-\sum_{r,s=1}^3\sum_{n,m}^{\infty}c_n^{(r)} \hat V_{nm}^{rs}b_m^{(s)}
\label{V3gh}
\ee
The dual vertex is 
\be
| V_{3}\rangle={\cal K}\,e^{E}|\hat \omega\rangle_{123}
\quad\quad
{E}=\sum_{r,s=1}^3\sum^\infty_{n,m}c_n^{(r)\,
\dagger} V_{nm}^{rs}b_m^{(s)\, \dagger}\label{V3gh'}
\ee

The range of $m,n$ is not specified. However,
for reasons that will become clear later, we would like to interpret 
the matrices $\hat V_{nm}^{rs}$  and 
$V_{nm}^{rs}$ as square long--legged matrices $(ll)$.
But, as soon as we try to evaluate, for instance,
contractions like $\langle \hat V_{3}|\omega\rangle_{123}$ in order to 
compute the constant ${\cal K}$, a problem arises linked to the 
presence in the exponent (\ref{V3gh}) of couples of conjugate
operators $c_0,b_0$, $c_{-1},b_1$ and $c_1, b_{-1}$. In order to appreciate 
this problem let us consider
the simple case of $e^{c_0V_{00}b_0}$. Interpreting this expression literally 
one gets
\be
e^{c_0V_{00}b_0}&=& 
1+ c_0 V_{00} b_0 + \frac 12 c_0 V_{00} b_0\,c_0 V_{00} b_0+\ldots\0\\
&=& 1+ c_0(V_{00}+ \frac 12 V_{00}^2+\ldots )b_0= 1+ c_0(e^{V_{00}}-1)b_0
\label{wrong} 
\ee
It follows that, when inserted in 
$\langle \hat V_{3}| \omega\rangle_{123}$ a term like 
this does not yield 1, as one would expect. Moreover
if, instead of the single zero mode we have considered here for simplicity, 
we had three, the result would be even more 
complicated.  All this is not natural. 
Let us recall that the meaning of $\hat V_{nm}^{rs}$ (see \cite{leclair} 
and below) is the coefficient of the monomial
$z^{m+1} w^{n-2}$ in the expansion of $\langle \hat V_3|R(c^{(s)}(z)
b^{(r)}(w))|\omega_{123}\rangle $ in powers of $z$ and $w$ (with
opposite sign). Therefore 
interpreting the exponentials in (\ref{V3gh}) as in (\ref{wrong}) 
is misleading. It is clear that what they really mean is something else. 
To adapt the oscillator formalism to the desired meaning we proceed as 
follows.

Let us introduce new conjugate operators $\eta_a, \xi^\dagger_a$, $a=-1,0,1$,
in addition to $c_n,b_m$, such that
\be
[\eta_a,\xi_b^\dagger]_+=\delta_{ab}\label{etaxi}
\ee
and they anticommute with all the other oscillators. Moreover we require them to 
satisfy
\be 
\eta_a|0\rangle =0 ,\quad\quad \langle 0 |\xi_a^\dagger =0\label{etaxi1}
\ee
while
\be
\langle 0|\eta_a \neq 0, \quad\quad \xi^\dagger_a |0\rangle \neq 0
\label{etaxi2}
\ee
Now let us replace in the exponent of (\ref{V3gh}) $c_a$  
 with $\eta_a$  (but not $c_a^\dagger$ in the exponent of (\ref{V3gh'})
with $\eta_a^\dagger$ ) and
$b_a^\dagger$ in the exponent of (\ref{V3gh'}) with $\xi_a^\dagger$ 
(but not $b_a$ in the exponent of (\ref{V3gh}) with $\xi_a$ -- in fact 
$c_a^\dagger$ and $b_a$ will not be needed). With these rules 
$\langle \hat V_{3}|\omega\rangle_{123}= {\cal K}$ straightforwardly.
The matrices $\hat V_{nm}^{rs}$ and $V_{nm}^{rs}$ are naturally square 
long legged. The 
interpretation of $\hat V_{nm}^{rs}$ as the negative coefficient of order 
$z^{m+1}$ and $w^{n-2}$ in the expansion of $\langle \hat V_3|R(c^{(s)}(z)
b^{(r)}(w))|\omega\rangle_{123}$ in powers of $z$ and $w$, remains valid
provided one replaces $b_{-1}^{(r)\,\dagger},b_{0}^{(r)\,\dagger}, 
b_{1}^{(r)\,\dagger}$ in $b^{(r)}(w)$ with
$\xi_{-1}^{(r)\,\dagger},\xi_{0}^{(r)\,\dagger}, \xi_{1}^{(r)\,\dagger}$.

We stress again that the substitution of $c_a$ with $\eta_a$ and
$b_a^\dagger$ with $\xi_a^\dagger$ is dictated by the requirement
of consistency of the interpretation of the Neumann coefficient
as expansion coefficients of the $b$-$c$ propagator.

\subsection{Ghost Neumann coefficients and their properties}

It is time to go to a concrete example. 
To this end one has to explicitly compute $\hat V_{nm}^{rs}$  
and $V_{nm}^{rs}$ in (\ref{V3gh},\ref{V3gh'}).
The method is well--known: one expresses the propagator with zero mode
insertions $\ll c(z) b(w)\gg$  in two different ways, first as a 
CFT correlator and then in terms of $\hat V_3$ and one equates the two 
expressions after mapping them to the disk via the maps
\be
f_i(z_i)=\alpha^{2-i} f(z_i) \, ,\, i=1,2,3\label{fi}
\ee
where
\be
f(z)=\Big{(} \frac{1+iz}{1-iz}\Big{)} ^{\frac{2}{3}}\label{f}
\ee
Here $\alpha=e^{\frac{2\pi i}{3}}$ is one of the three third roots of
unity. However this recipe leaves several uncertainties due especially
to the ghost insertions. For concreteness in Appendix A we make a specific 
choice of these insertions, in a way the simplest one: we set the insertions at 
infinity. Even so there remain some uncertainties which 
we fix by requiring certain properties, in particular cyclicity,
consistency with the $bpz$ operation and commutativity of the twisted 
matrices of Neumann coefficients (the motivation for the latter will become
clear further on). With this (arbitrary) choice, the ghost Neumann coefficients 
worked out in Appendix A satisfy the following set of properties:

\begin{itemize}
\item{\it cyclicity}

\be
\hat V_{nm}^{rs}=\hat V_{nm}^{r+1,s+1},\label{cyclgh}
\ee
 
\item{\it $bpz$ consistency}

\be
(-1)^{n+m} V_{nm}^{rs}= \hat V_{nm}^{rs}\label{bpz}
\ee

\item{\it commutativity} 

Its meaning is the following. Defining 
$X=\hat CV^{rr}$, $X^+=\hat C V^{12},X^-= \hat CV^{21}$, we have
\be
X^{rs}X^{r's'}= X^{r's'}X^{rs}\label{commut}
\ee
for all $r,s,r',s'$. In addition we have
\be
X+X^++X^-=1\label{fund1}
\ee
and
\be
X^+X^- = X^2 -X,\quad\quad X^2 +(X^+)^2+(X^-)^2=1\label{fund2}
\ee
It should be stressed that all the $X^{rs}$ matrices are $(ll)$.
\end{itemize}

\subsection{Formulas for wedge states}

Our next goal is to define recursion relations for the
ghost wedge states. To start with we define the star product of squeezed 
ghost states of the form
\be
|S\rangle={\cal N} 
\exp \left(c^\dagger S b^\dagger\right)|0\rangle\label{squeezed}
\ee
We notice that since the vacuum is $|0\rangle$ we are implicitly
referring to the natural normal ordering.
The star product of two such states $|S_1\rangle$ and $|S_2\rangle$
is the $bpz$ of the state
\be
\langle \hat V_3||S_1\rangle_1 |S_2\rangle_2\label{starpr}
\ee
However this formula needs some specifications. We remark that 
the problem pointed out above, linked to the presence of couples 
of conjugate oscillators in the exponents, is present both in 
(\ref{squeezed}) and (\ref{starpr}). We solve it
as we did in section 2,  with the help of additional oscillators 
$\eta_a, \xi_b^\dagger$. We interpret  for instance
(\ref{squeezed}) as follows. We replace the new oscillators 
in it as in section 2,
then we exploit the anticommutativity properties of the latter to move 
them to the right and apply them to $|0\rangle$, then
we substitute back $b_a^\dagger$ in the place of $\xi_a^\dagger$.
 The upshot of this operation is that no $b_a^\dagger$
oscillator will survive and the state (\ref{squeezed}) takes the form
\be
|S\rangle={\cal N} 
\exp \left(\sum_{n=-1}\sum_{m=2}c_n^\dagger S_{nm} b_m^\dagger\right)
|0\rangle
\label{squeez}
\ee
That is, the matrix $S_{nm}$ in the exponent is lame $ls$. This is
the precise meaning we attach to (\ref{squeezed}). 
Let us notice that the $bpz$ dual expression of (\ref{squeez}) is
\be
\langle S|={\cal N}\langle 0| 
\exp \left(-c \hat C S\hat C b\right)\label{squeezeddual}
\ee
The matrix $S$ here is $ll$.

After this specification let us define the star product 
of $|S_1\rangle$ and $|S_2\rangle$. Let us recall the three strings 
vertex (\ref{V3gh},\ref{V3gh'}). Remembering
the discussion before  (\ref{squeez}) we conclude that
$\hat V_{nm}^{rs}$ is $sl$ for $r=1,2$ and $ll$ for $r=3$, while
$V_{nm}^{rs}$ is $ls$ for $r=1,2$ and $ll$ for $r=3$.

In evaluating this product we will have to evaluate vev's of the type
\be
\langle\hat {0}|\exp \Big({cFb+c\mu+\lambda b}\Big)\,
\exp\Big({c^{\dagger}Gb^{\dagger}
+\theta b^{\dagger}+c^{\dagger}\zeta}\Big)|{0}\rangle\label{trial}
\ee
Here we are using an obvious compact notation: $F,G$ denotes matrices
$F_{nm}, G_{nm}$, and $\lambda,\mu,\theta,\zeta$ are anticommuting
vectors $\lambda_n,\mu_n,\theta_n,\zeta_n$. We expect the result of this 
evaluation to be   
\be
&\langle\hat{0}|\exp\left({cFb+c\mu+\lambda b}\right)
\exp\left({c^{\dagger}Gb^{\dagger}
+\theta b^{\dagger}+c^{\dagger}\zeta}\right)|{0}\rangle&\nonumber\\
&& \0 \\
&=\det(1+FG)\,\exp\left({-\theta\frac{1}{1+FG}F\zeta-\lambda\frac{1}{1+
GF}G\mu-\theta\frac{1}{1+FG}\mu+\lambda\frac{1}{1+GF}\zeta}\right)&
\label{formula}
\ee

In order for this formula to hold in (\ref{trial}) the operator denoted 
$b,c$ must be creation operators with respect to $\langle \hat 0|$ and 
annihilation operators with respect to the $|0\rangle$ vacuum. Viceversa
the oscillators denoted $c^\dagger, b^\dagger$ must be all creation
operator with respect to $|0\rangle$, and annihilation operators
with respect to $\langle \hat 0|$. But this is precisely what happens 
if we assume the definition (\ref{squeez}) for the squeezed states and
(\ref{V3gh}) for the vertex with the summation over $n$ starting from 2 
(which is consistent with the interpretation by means of $\xi_a^\dagger$ and
$\eta_a$, as before (\ref{squeez})). 

Therefore it is correct to use formulas like (\ref{formula}) in order to
evaluate the star product (\ref{starpr}), but in this case the matrices $F$ 
and $G$ will be lame ($ls$ or $sl$ as the case be), while analogous 
considerations apply to the vectors $\lambda,\mu,\theta,\zeta$ 
($\lambda,\zeta$ 
are long vectors, while $\mu, \theta$ are short).
The star product of two squeezed states like (\ref{squeezed}) is
\be
|S_1\rangle \, \star \, |S_2 \rangle = |S_{12}\rangle\0
\ee
where the state in the RHS has the same form as (\ref{squeezed}), with the matrix $S$
replaced by $S_{12}=\hat CT_{12}$. The latter is given by the familiar formula
\be
T_{12}= X + (X^+,X^-) \,\frac 1{1- \Sigma_{12} {\cal V}} \,\Sigma_{12} 
\,\left(\begin{matrix} X^-\\ X^+ \end{matrix}\right)\label{T12}
\ee
where
\be
\Sigma_{12} =\left(\begin{matrix}\hat CS_1&0\\0&\hat CS_2\end{matrix}\right),
\quad\quad
{\cal V}= \left(\begin{matrix}X & X^+\\X^-& X\end{matrix}\right) 
\label{SigmaV}
\ee 
The normalization of $|S_{12}\rangle$ is given by
\be
{\cal N}_{12}={\cal N}_1\, {\cal N}_2 \,
\det \left(1 -{\cal V} \Sigma_{12}\right)\label{norm12}
\ee
Notice that in this formula the four matrices in ${\cal V} \Sigma_{12}$ are 
$ss$. 

These expressions are well defined. However, since they are expressed in 
terms of lame matrices we cannot operate with them in the same way we 
usually do with the analogous matrices of the matter sector. For that 
one needs the identities proved in the previous section, which are only valid
for long square matrices.  Luckily
in the case of the  wedge states it is possible to overcome this difficulty.

When computing a star product we would like to be able to apply 
the formulas of subsection 2.1, which are expressed in terms 
of long square matrices. 
To this end we would like (\ref{formula}) to be expressed in terms of long 
square matrices, rather than of lame matrices. This is possible at the price 
of some modifications.

Let us introduce the new conjugate operators $\eta_a, \xi^\dagger_a$, 
$a=-1,0,1$, as above, see (\ref{etaxi},\ref{etaxi1},\ref{etaxi2}) and
let us replace in (\ref{formula}) $c_a$ (but not $c_a^\dagger$) 
with $\eta_a$ and $b_a^\dagger$ (but not $b_a$) with $\xi_a^\dagger$. 
Then in the RHS long square matrices and long vectors will feature 
(instead of lame matrices and short or long vectors). 
In the sequel we will use (\ref{formula}) in this sense. Such 
modifications of course are not for free. We have to justify them
\footnote{In the previous cases the introduction of the new oscillators
was simply an auxiliary tool to help us interpret such formulas as 
(\ref{squeez}). We could have done without them by {\it ad hoc} definitions.
But now we are tampering with vev's, therefore we have to make sure
that we do not modify anything essential.}. 
We will show later on that in the case of the wedge states
such a move is justified.

Once this is done the calculation of the star product works smoothly  without
any substantial difference with respect to the matter case. 
The formulas are the same eqs.(\ref{T12},\ref{SigmaV}) and (\ref{norm12}) 
above, but expressed in terms of long square matrices to which we can apply 
the identities of subsection 2.1.  
This allows us to treat the ghost squeezed states in a way completely similar to the 
matter squeezed states. Of course it remains for us to comply with the 
promise we made of showing that we are allowed to use long square matrices.

The wedge states are now defined to be squeezed states 
$|n\rangle\equiv |S_n\rangle$ 
that satisfy the recursive star product formula
\be
|n\rangle \star |m\rangle = |n+m-1\rangle\label{wedgenm}
\ee
This implies that the relevant matrices $T_n=\hat C S_n$ satisfy the recursion 
relation
\be
T_{n+m-1}= \frac {X-(T_n+T_m)X+ T_nT_m}{1- (T_n+T_m)X+T_nT_mX}\label{recur1}
\ee
or
\be
T_{n+1}= X\frac {1-T_n}{1-T_nX},\label{recur}
\ee
and the normalization constants are given by
\be
{\cal N}_{n+1}={\cal N}_n\, {\cal K} \, \det \left(1-T_n X\right)  
\label{normrecur}
\ee
These relations are derived 
under the hypothesis that $T_n$ and $X, X^+,X^-$ commute and
by using the identities of subsection 2.1. The solution to 
(\ref{recur}) is well--known, \cite{Furu,Kishimoto}. We repeat the derivation
in order to stress its uniqueness.  
We require
that $|2\rangle$ coincide with the vacuum $|0\rangle$, both for the matter 
and the 
ghost sector\footnote{It is worth recalling that our purpose in this paper 
is to complete the proof started in  \cite{BMST} of (\ref{ghostwedge})
\be
e^{ - \frac {n-2}2\left({\cal L}^{(g)}_0 +
{\cal L}_0^{(g)\dagger}\right)}|0\rangle
= {\cal N}_n\, e^{ c^\dagger S_n b^\dagger}|0\rangle\equiv |n\rangle 
\0
\ee
that is, that the LHS does represent the ghost wedge states. In this light 
the requirement that the wedge state $|n\rangle$ with $n=2$ coincide with 
the vacuum state is natural.}. 

This implies in particular that $T_2=0$ and ${\cal N}_2=1$, which
entails from (\ref{recur}) that $T_3=X$, $T_4= \frac X{1+X}$, etc. That is $T_n$
is a uniquely defined function of $X$. But $X$ can be uniquely expressed in terms of 
the sliver matrix $T$
\be
X=\frac T{T^2-T+1}\label{XT}
\ee
a formula whose inverse is well--known, \cite{KP,RSZ2}
\be
T= \frac 1{2X} \left(1+X- \sqrt{(1-X)(1+3X)}\right)\label{sliver}
\ee
Therefore $T_n$ can be expressed as a uniquely defined function of $T$.
Now consider the formula
\be
T_n= \frac {T+(-T)^{n-1}}{1-(-T)^n}\0
\ee
It satisfies (\ref{recur}) as well as the condition $T_2=0$, therefore 
it is the unique solution to (\ref{recur}) we were looking for. 

So far the states $|n\rangle$ have been defined solely in terms of the 
three strings vertex. One might ask what is their connection with
the wedge states defined as surface states, \cite{GRSZ2,Schnabl2,FKM,Schnabl05}. This connection 
can be established: it can be shown that, with the appropriate insertion
of the zero modes, the surface wedge 
matrix $S_3$ is actually $V^{rr}$, i.e. $T_3=X$. 

It is simple to see that similarly (\ref{normrecur}) has a unique 
solution satisfying ${\cal N}_2=1$ and ${\cal K}={\cal N}_3$.

\subsection{Commutation relations with $K_1$}

What we have done so far is all very good, but the concrete example of vertex
constructed in Appendix A is only academical, as the 
following remark shows.
In \cite{BMST} we diagonalized the LHS of \ref{ghostwedge} on the basis of
weight 2 differentials, 
in which the operator $K_1$ is diagonal. In order to be able to compare
this result with the wedge states defined above we have to make sure that
also the matrices $T_n, X,X_+,X_-$ can be diagonalized in the same basis.
In this subsection we will discuss this problem.  

Let us recall the definition of $K_1$:
\be
K_1= \sum_{p,q\geq -1} c_p^\dagger\, G_{pq} \,b_q + \sum_{p,q\geq 2}
b_{\bar p}^\dagger\,H_{\bar p\bar q}\,
c_{\bar q}- 3c_2\,b_{-1}\label{K1gh}
\ee
where
\be
&&G_{pq}= (p-1) \delta_{p+1,q} + (p+1)\delta_{p-1,q},\0\\
&&H_{\bar p\bar q}= (\bar p+2) \delta_{\bar p+1,\bar q}
+ (\bar p-2)\delta_{\bar p-1,\bar q}\label{GF}
\ee
$G$ is a square long--legged matrix and $H$ a square
short--legged one. In the common overlap we have $G=H^T$.
We notice immediately that $K_1$ annihilates the vacuum
\be
K_1|0\rangle=0\label{K10}
\ee
What is important for us is that the action of $K_1$ commutes with 
the matrices we want to diagonalize.
Now let $T_n=\hat C S_n$,
where $S_n$ is the matrix of the squeezed state representing $|n\rangle$.
We have seen that $T_n$ can be either lame or square $(ll)$.
Since we want to diagonalize (\ref{recur}) we must consider the second 
alternative. But in order to arrive at square $(ll)$ matrices, at the beginning 
of this section we introduced into the game the conjugate oscillators
$\eta_a,\xi_a^\dagger$, $a=-1,0,1$. Therefore, to be consistent,
they must appear also in the oscillator representation of $K_1$.
This can be done as follows. 
 
We write down $K_1$ as
\be
K_1= \sum_{p,q\geq -1} c_p^\dagger\, G_{pq} \,b_q + \sum_{p,q\geq -1}
b_{p}^\dagger\,H_{ p q}\,c_{q}\label{K1gh'}
\ee
where $G$ and $H$ have the same expression as before, but now also
$H$ is square long legged and $H=G^T$. What is important is
that in the expression $b^\dagger Hc$ we understand that $b_a^\dagger$
is replaced by $\xi_a^\dagger$ and $c_a$ is replaced by $\eta_a$ (for
simplicity we dispense with writing the new $K_1$ explicitly). 
If we write
\be
&& b(z) = \sum_{n\geq 2} b_n z^{-n-2} + \sum_{-1\leq a\leq 1} \xi_a^\dagger
z^{-a-2} +\sum_{n\geq 2} b_n^\dagger z^{n-2}\label{bz}\\
&& c(z) = \sum_{n\geq 2} c_n z^{n-1} + \sum_{-1\leq a\leq 1} \eta_a
z^{-a+1} +\sum_{n\geq 2} c_n^\dagger z^{n+1}\label{cz}
\ee
we find the expected conformal action of $K_1$ on these fields. For instance
\be
[K_1,b(z)]&=& - \sum_n \left((n-1) \,b_{n+1} +(n+1)b_{n-1}\right) 
\, z^{-n-2}\0\\
&=& (1+z^2)\partial b(z) +4 z\,b(z)\0
\ee
after replacing back $b_a^\dagger$ for $\xi_a^\dagger$.
 
On the basis of this discussion we expect therefore that
\be
[G,T_n]=0\label{GTn}
\ee
as square long legged matrices. In particular we should find that
$G$ commute with $X$. One can however show that this is not the case for
the vertex explicitly constructed in Appendix A.
Therefore that vertex has many good properties but not this one.

However we will show below it is very plausible that a three strings vertex
that satisfies also (\ref{GTn}) exists. Therefore in the sequel we 
imagine that we have done everything with this vertex and will try to
justify its existence {\it a posteriori}.

\section{The diagonal recursive relations for wedge states}

So far we have worked, so to speak, on the RHS of eq.(\ref{ghostwedge}). 
It is now time to make a comparison with the left hand side.
In \cite{BMST} we showed that
\be
e^{ - \frac {n-2}2\left({\cal L}^{(g)}_0 +
{\cal L}_0^{(g)\dagger}\right)}|0\rangle= 
e^{\eta (t)} e^{c^\dagger \alpha(t) b^\dagger}|0\rangle\label{ghwed1}
\ee
where $t= (2-n)/2$,
\be
\alpha(t) = A \frac {{\rm sinh}\left(\sqrt{(D^T)^2-BA}\,t\right)}
{ \sqrt{(D^T)^2-BA}\, {\rm cosh} \left(\sqrt{(D^T)^2-BA}\,t\right)
-D^T\, {\rm sinh}\left(\sqrt{(D^T)^2-BA}\,t\right)}
\label{alphat}
\ee
and
\be
\eta(t) = - \int_0^{t}\,dt'\,{\rm tr} (B\alpha(t'))\label{etangh}
\ee
and $A,B, D^T$ are matrices extracted from ${\cal L}^{(g)}_0 
+{\cal L}_0^{(g)\dagger}$.
In particular $D^T$ as well as the combination $(D^T)^2-BA$ are $(ss)$ matrices,
while $A$ is lame $(ls)$. The purpose of the paper was to show that the RHS of
(\ref{alphat}), multiplied by the twist matrix $\hat C$, does satisfy the recursion 
relations (\ref{recur}) and (\ref{normrecur}).
This was achieved by diagonalizing the matrices $\tilde A=\hat C A,D^T$ and 
$(D^T)^2-BA$ on the 
weight 2 basis $V_n^{(2)}(\kappa)$, with $n=2,3,\ldots$. We concluded that if we are
allowed to replace in (\ref{recur},\ref{normrecur}) the matrices by their eigenvalues 
in such a basis, the recursion relations can be shown to be true. What remained to be 
proved was precisely the correctness of replacing in such formulas the matrices by 
their eigenvalues. We are now in the position to do it.

Let us examine first (\ref{alphat}), multiplied from the left by the twist 
matrix. The RHS is the product of $\tilde A$ by a matrix which is
diagonal in the weight 2 basis and is of type $(ss)$. Therefore when we apply 
the latter to a vector $V^{(2)}_s$ with components 
$(V_2^{(2)}(\kappa),V_3^{(2)}(\kappa),\ldots  )$, we obtain
the same vector multiplied by the eigenvalue. When we next apply 
$\tilde A$ from the left to the resulting vector, things are a little bit 
more complicated because $A$ is an $(ls)$ matrix. The vector ensuing from 
the operation would seem to have three additional 
entries with $n=-1,0,1$, therefore making meaningless even the idea of 
eigenvalue and 
eigenvector. However it was shown in \cite{BMST} that, 
\be
&&\sum_{q=2}^\infty \tilde A_{p q} \, V_{q}^{(2)}(\kappa) = 
{\mathfrak a}(\kappa) 
V_{p}^{(2)}(\kappa), \quad\quad p=2,3,\ldots\label{eigenforA}\\
&&\sum_{q=2}^\infty \tilde A_{a q}\, V_{q}^{(2)}(\kappa) =0,\quad\quad 
a=-1,0,1\label{eigenA=0}
\ee
(for a new proof of (\ref{eigenA=0}) see Appendix C).
I.e., not only is the $(ss)$ submatrix of $\tilde A$ diagonal in the 
weight 2 basis with eigenvalue ${\mathfrak a}(\kappa)$, but the potential 
additional vector elements vanish. 
This allows us to conclude that the same property is shared by the matrix 
$\alpha(t)$. That is, when applying $\alpha(t)$ to the weight 2 basis vector 
$V_s^{(2)}$ as above, we obtain the 
corresponding eigenvalue multiplying the same vector (without 
additional components): $\alpha(t) V_s^{(2)} = \alpha(\kappa,t) V_s^{(2)}$.

Now let us apply the above to (\ref{recur}). The latter is formulated in 
terms of long square matrices whose $(ls)$ part has the form $\alpha(t)$. 
Now we can interpret 
(\ref{recur}) as an infinite series expansion, in which each term is a 
monomial of (possibly different) matrices whose $(ls)$ part has the form 
$\alpha(t)$. Let us consider the weight 2 basis vector $V_s^{(2)}$ 
extended by adding three 0 components in position $-1,0,1$, and let us call
it $V_l^{(2)}$. When we apply any of the above matrices to it, we get the 
same extended vector multiplied by the matrix eigenvalue: for instance
$\alpha(t) V_l^{(2)} = \alpha(\kappa,t) V_l^{(2)}$. Therefore we can 
repeat the operation as many times
as needed for any monomial and obtain the same vector multiplied by the 
monomial in which each matrix is replaced by its eigenvalue. Re-summing 
the series we obtain that the relation (\ref{recur}) applied to the 
weight 2 basis vector becomes a relation of the same 
form with the matrices replaced by the corresponding eigenvalues. 
But in \cite{BMST} we checked that this relation for the eigenvalues is true. 
This, which is the main argument in \cite{BMST} and the present paper, 
and is intended to lead to the proof of  (\ref{ghostwedge}),
has been laid out so far in a rather patchy way due to its complexity.  
For the sake of clarity it is worth reviewing it in full, even at the price 
of some repetitions. 

\subsection{Proof of the diagonal recursive relations for wedge states}

We wish to show 
that the eigenvalues of the matrices $S_n$ in (\ref{ghostwedge}) satisfy 
(\ref{recur}) and (\ref{normrecur}). 
This sentence has to be unambiguously understood. First we notice that
$T_2=0$, which is consistent with $|2\rangle$ being identified with the
vacuum $|0\rangle$ and ${\cal N}_2=1$. Now proving  (\ref{recur})
means proving two things:
\be
T_3=X \label{firsteq}
\ee
and
\be
T_{n+1}= T_3\frac {1-T_n}{1-T_n T_3},\label{secondeq}
\ee
This second equation is demonstrated by setting 
\be
{T_n}\equiv \hat C S_n = \tilde \alpha\left(-\frac{n-2}2\right)\label{idengh}
\ee
and using $\tilde \alpha$ given by eq.(\ref{alphat}). This gives
the explicit expression
\be
{T_n}= -\tilde A
\frac {{\rm sinh}\left(\sqrt{\Delta}\,\frac{n-2}2\right)}
{ \sqrt{\Delta}\, {\rm cosh} \left(\sqrt{\Delta}\,\frac{n-2}2\right)
+D^T\, {\rm sinh}\left(\sqrt{\Delta}\,\frac{n-2}2\right)}\label{Tngh}
\ee
where $\Delta =(D^T)^2-BA$. On the basis of the 
remarks made at the beginning of this section, we replace everywhere the 
matrices by their eigenvalues
\be
\sqrt{\Delta} =\frac {\pi |\kappa|}2,\quad\quad
\tilde A= \frac {\kappa\pi}{2\, {\rm sinh}\left(\frac {\kappa\pi}2\right)},\quad\quad
D^T= \frac {\kappa\pi}2 \, {\rm coth}\left(\frac {\kappa\pi}2\right)\label{A&Cgh}
\ee
By inserting (\ref{idengh}) into (\ref{secondeq}) one can see that the latter
is satisfied (see section 2.5 of \cite{BMST} for details) if
\be
D^T+\tilde A= \sqrt{\Delta}\, {\rm coth} \left(\frac {\kappa\pi}2\right)\0
\ee
This is immediately verified using (\ref{A&Cgh}).

Next, in order to prove (\ref{firsteq}), we recall that (\ref{secondeq})
can be solved by
\be
T_n= \frac {T+(-T)^{n-1}}{1-(-T)^n}\label{recursolution}
\ee
for some matrix $T$. This matrix is easily identified to be 
$T\equiv T_\infty$ (this makes sense because the absolute value (of the
eigenvalue) of $T$ turns out to be $<1$:
$T= -e^{-\frac {|\kappa|\pi}2}$). But, from the defining
eq.(\ref{wedgenm}), $T$ represents the sliver \cite{KP,RSZ2,RSZ3}. 
Therefore it is related to $X$ by eq.(\ref{sliver}) or by its inverse
\be
X= \frac T{T^2-T+1}\label{inversesliver}
\ee
This is precisely (\ref{recursolution}) for $n=3$. Therefore (\ref{firsteq})
is satisfied and in addition this tells us that the eigenvalue
of $X$ is
\be
X= - \frac 1{1+ 2 {\rm cosh}\left(\frac {\kappa\pi}2\right)}\label{eigenX}
\ee
Since the recursive constraints propagates this identification to all
the wedge states this complete our proof\footnote{It might seem at first sight
that eq.(\ref{eigenX}) contradicts the well--known formula found by Gross
and Jevicki, \cite{GJ1,GJ2,MM}, 
\be
X=- E\frac {M}{1+2M} E^{-1}\label{XM}
\ee
which relates the twisted ghost Neumann coefficients matrix $X$ with the
corresponding matter matrix of Neumann coefficients $M$.  
If we naively diagonalize $X$ and $M$ on the matter basis of 
eigenvectors and use the result of \cite{RSZ1}, we obtain a value for $X$ 
different from (\ref{eigenX}). 
However this is an `optical' effect: eq.(\ref{XM}) is certainly true 
numerically, but 
$X$ and $M$ act on different spaces, therefore they are different operators.
Each must be diagonalized in its own space. They cannot be diagonalized
using the same basis. There is thus no room for
contradiction between (\ref{XM}) and (\ref{eigenX}).}.

Let us come to the normalization constants ${\cal N}_n$. They must  
satisfy a recursion relation
\be
{\cal N}_n\, {\cal K} \, \det \left(1-T_n X\right) = {\cal N}_{n+1}
\label{normrecurgh}
\ee
where ${\cal K}$ is some constant to be determined. We fix it by requiring
that ${\cal N}_2=1$ so that the wedge state $|2\rangle$ coincides with
the vacuum $|0\rangle$.
We have
\be
\eta_n = - \int_0^{t_n}\,dt\,{\rm tr} (B\alpha)\,=\,
-\int_0^{t_n}\,dt\,{\rm tr} (\tilde A \tilde \alpha)\label{etangh'}
\ee
where the trace is over the weight 2 basis. Now identifying
\be
{\cal N}_n = e^{\eta_n}, \0\label{etanNngh}
\ee
plugging in the relevant eigenvalues and proceeding as in section (2.5)
of \cite{BMST} one can easily verify that (\ref{normrecurgh}) is 
satisfied.   

This completes our proof that the squeezed states in the
midterm of (\ref{ghostwedge}) have the same eigenvalue as the 
ghost wedge states in the oscillator formalism. 

To complete this argument we must show that our choice of enlarging the Fock 
space at the beginning of this section is justified in the case of the 
wedge states. Since this requires the same type of arguments as  
in the previous subsection 
and is somewhat repetitious, we will account for it in Appendix B.

Finally let us remark that without the commutativity property of the
twisted Neumann coefficients matrices spelled out in section 2,
it would be impossible to reproduce the results of \cite{BMST} where
the matrices $A,B,C,D^T$ commute (in the appropriate way).

The results we have obtained in this section consolidates the result
obtained in \cite{BMST}, however is not yet the end our proof of 
(\ref{ghostwedge}). In the next section we explain why.

\section{Matrix reconstruction from the spectrum}

So far our argument has been carried out by replacing the matrices involved
with their eigenvalues. It would seem that we are done with the proof of
(\ref{ghostwedge}). However what we have to show is not only that the 
eigenvalues of the matrices featuring in the RHS of eq.(\ref{ghwed1})
coincide with the eigenvalues of the matrix $S_n$ in (\ref{squeezed}),
where $S_n$ satisfies the recursion relation (\ref{recur}), but {\it that the 
matrices themselves coincide}. Now, in general, if one knows eigenvalues 
and eigenvectors 
of a matrix operator one can reconstruct the original matrix. This is true
for the matter sector of (\ref{ghostwedge}), but in the ghost sector this
is not the case. In the ghost sectors things are unfortunately more 
complicated due to the existence of zero modes. This section is devoted to 
explaining this additional complication.

So far our argument has
consisted in applying the matrices involved such as $\tilde A, D^T$ and in 
particular $\alpha(t)$ to the weight 2 basis vector $V^{(2)}$. As shown in 
\cite{BMST}, the exponent $c^\dagger \alpha b^\dagger$ in (\ref{ghwed1}) 
can be written as follows  
\be
c^\dagger \alpha b^\dagger &=& \sum_{n=-1, m=2} c^\dagger_n \alpha_{nm}(t) 
b_m^\dagger = \sum_{n=-1, m=2}^\infty \int d\k \, d\k'\,\tilde c^\dagger(\k) 
\tilde V_n^{(-1)}(\k) \tilde \alpha_{nm}(t)  \tilde V_m^{(2)}(\k')\, 
b^\dagger(\k')\0\\
&=& \sum_{n=2}^\infty \int d\k\, d\k'\, 
\tilde c^\dagger(\k)\tilde V_n^{(-1)}(\k)
\tilde \alpha(\k,t) \tilde V_n^{(2)}(\k')b^\dagger(\k')=
\int d\k\,\tilde c^\dagger(\k)\tilde \alpha(\k,t)\, 
b^\dagger(\k)\label{cABdagger}
\ee
where we have introduced
\be
(-1)^n\, c_n^\dagger = \int d\k \,\tilde c^\dagger(\k) \,\tilde V_n^{(-1)}(\k),
\quad\quad
b_n^\dagger = \int d\k\, b^\dagger(\k)\,\tilde V_n^{(2)}(\k),\quad n\geq 2
\label{c(k)}
\ee

It is clear that if the LHS of (\ref{cABdagger}) is exactly equal to the RHS,
our proof is complete. However the question is: in
(\ref{cABdagger}) we went from left to right, i.e. from the LHS we derived 
the RHS. Can we go the other way?
In other words, given that we know the eigenvalue of some matrix in the
weight 2 basis (or, for that matter, in the weight -1 basis) can we 
reconstruct
the original matrix? For instance, we notice that in the intermediate steps of 
(\ref{cABdagger}) the summation over $n=-1,0,1$ has disappeared. The 
obvious question is: how can we reconstruct these modes when we run the argument
from right to left?  

To start with let us recall the definitions of the two bases.
The unnormalized basis (weight 2 basis) is given by
\be
f^{(2)}_\k(z) = \sum_{n=2} V_n^{(2)}(\k) \, z^{n-2}\label{bbasis}
\ee
in terms of the generating function
\be
f^{(2)}_\k(z) = \left(\frac 1{1+z^2}\right)^2 \, e^{\k \arctan (z)}=
1+\k z +\left(\frac {\k^2}2 -2\right) z^2+\ldots\label{genf}
\ee
Following \cite{Belov1,Belov2}, (see also Appendix B of \cite{BMST}), we normalize
the eigenfunctions as follows
\be
\tilde V_n^{(2)}(\k)= {\sqrt{A_2(\k)}} V_n^{(2)}(\k)
\label{nbbasis}
\ee
where
\be
 A_2(\k) = \frac {\k (\k^2+4)}
{2\sinh \left(\frac {\pi \k}2\right)}\0
\ee

The unnormalized weight -1 basis is given by
\be
f_\k^{(-1)}(z) = \sum_{n=-1} V_n^{(-1)}(\k) \, z^{n+1}\label{cbasis}
\ee
in terms of the generating function
\be
f^{(-1)}_\k(z) = (1+z^2) \, e^{\k \arctan (z)}=
1+\k z +\left(\frac {\k^2}2 +1\right) z^2+\ldots\label{f-1}
\ee
The normalized one is
\be
\tilde V_n^{(-1)}(\k)=\sqrt{A_{-1}(\k)} V_n^{(-1)}(\k),
\quad\quad \sqrt{A_{-1}(\k)} =  {\cal P} \frac 1{\k}
\frac {\sqrt{A_2(\k)}}{\k^2+4}\label{normV-1n}
\ee
where ${\cal P}$ denotes the principal value. We reported in \cite{BMST}
the biorthogonality
\be
\int_{-\infty}^{\infty} d\k\, \tilde V^{(-1)}_{n}(\k)\,\tilde
V^{(2)}_{m}(\k)= \delta_{n,m}, \quad\quad n\geq 2\label{biorthog}
\ee
and `bi--completeness' relation
\be
\sum_{n=2}^\infty \, \tilde V^{(-1)}_{n}(\k)\, \tilde V^{(2)}_{n}(\k')=
\delta(\k,\k')\label{bicompleteness}
\ee
taking them from \cite{BeLove}. These relations can be formally proved, but 
it is evident that they have to be handled with care. Let us 
recall again eqs.(\ref{eigenforA}) and (\ref{eigenA=0}), 
which turned out to be crucial in the previous sections, and let us do 
the following. We multiply
(\ref{eigenA=0}) by $V^{(2)}_n(\k)$ and integrate over $\k$: we get 
$A_{an}=0$ for $n\geq 2$, which is evidently false. On the other hand
eq.(\ref{eigenA=0}) is correct (we present a new demonstration of it in 
Appendix C). Therefore it is apparent that in the above exercise
we did something illegal. This can only be the exchange between the 
(infinite) summation and the integration over $\k$. We remark that 
the same kind of exchange occurs also in the intermediate steps of 
(\ref{cABdagger}). We are therefore warned that in doing so we may lose 
some information. The question is: is there a way to repair the illegality
we commit in this way and recover the full relevant information? 

In mathematical terms this involves the problem of the spectral 
representation for lame operators. Unfortunately we have not been 
able to find any treatment of this problem in the mathematical literature.
We proceed therefore in a heuristic way.

\subsection{The problem}

Let us analyze the reconstruction of the matrix $\tilde A$. 
Since $\sum_{l=2}^\infty \tilde A_{nl} \tilde V_l^{(2)}(\k) = 
{\mathfrak a}(\k)\tilde V_n^{(2)}(\k)$,
we might argue as follows
\be
\int_{-\infty}^{\infty} d\k\tilde V_m^{(-1)} (\k){\mathfrak a}(\k)\tilde V_n^{(2)}(\k)
= \sum_{l=2}^\infty \tilde A_{nl} \int_{-\infty}^{\infty} \tilde V_m^{(-1)}(\k) 
V_l^{(2)}(\k)= \tilde A_{nm}
\ee
using the bi-orthogonality relations (\ref{biorthog}). 
Therefore we should be able to reconstruct the $\tilde A$ matrix starting from
\be
{\mathfrak a}(\k) = \frac {\pi\k}2 \frac 1{\sinh \left( \frac {\pi\k}2 \right)}\0
\ee
and the bases. Here are the first few basis elements
\be
&&V_2^{(2)}(\k) =1,\quad\quad V_3^{(2)}(\k)=\k, \quad\quad  
V_4^{(2)}(\k)=\frac {\k^2-4}2\label{b2}\\
&&V_5^{(2)}(\k)=\frac 16 \k (\k^2-14),\quad\quad V_6^{(2)}(\k)=
\frac 1{24} (\k^4-32\k^2+72),\quad\ldots\0
\ee
and 
\be
&&V_2^{(-1)}(\k) =\frac 16 \k(\k^2+4),\quad\quad V_3^{(-1)}(\k)=\frac {\k^2(\k^2+4)}{24},
\0\\ 
&&  V_4^{(-1)}(\k)=\frac \k{120} (\k^4-16)\quad\quad
 V_5^{(-1)}(\k)=\frac {\k^2}{720} (\k^4-10\k^2-56)   \label{b-1}\\\
&&V_6^{(-1)}(\k)=\frac {\k}{5040} (\k^6-28\k^4-56 \k^2+288),\quad \ldots\0
\ee
These must be multiplied by the normalization factor
\be
\sqrt{A_2(\k)} = \sqrt{\frac {\k(\k^2+4)}{2 \sinh\left( \frac {\pi\k}2 \right) }}\0
\ee
and
\be
\sqrt{A_{-1}(\k)}=  {\cal P} \frac 1{\k} 
\sqrt{\frac \k{2(\k^2+4)\sinh\left( \frac {\pi\k}2 \right)}} \0
\ee
so that
\be
\sqrt{A_2(\k) \,A_{-1}(\k)}= \frac 1{2 \sinh\left( \frac {\pi\k}2 \right) }\0
\ee
Using these formulas we get
\be
\tilde A_{22}\,&=&\, \frac \pi{24} \int_{-\infty}^{\infty} d\k\,
 \frac{\k^2(\k^2+4)}{\left(
\sinh\left( \frac {\pi\k}2 \right)\right)^2} \approx 0.5332\label{A22}\\
 \tilde A_{24}\,&=&\, \frac \pi{480} \int_{-\infty}^{\infty} d\k\, 
\frac{\k^2(\k^4-16)}{\left(
\sinh\left( \frac {\pi\k}2 \right)\right)^2} \approx -0.0762\label{A24}\\
\tilde A_{33}\,&=&\, \frac \pi{96} \int_{-\infty}^{\infty} d\k\, 
\frac{\k^4(\k^2+4)}{\left(
\sinh\left( \frac {\pi\k}2 \right)\right)^2} \approx 0.1524\label{A33}\\
\tilde A_{35}\,&=&\, \frac \pi{2880} \int_{-\infty}^{\infty} d\k\, 
\frac{\k^4(\k^2-14)(\k^2+4)}{\left(
\sinh\left( \frac {\pi\k}2 \right)\right)^2} \approx -0.0508\label{A35}
\ee
From the definition of the matrix $A_{nm}$ one shoulf get instead
\be
&&\tilde A_{22} = -\frac {12}{15}=-0.8,\quad\quad  
\tilde A_{24} = \frac {16}{35}\approx 0.457\0\\
&& \tilde A_{33} = -\frac {18}{35}\approx -0.514,\quad\quad
\tilde A_{35} = \frac {22}{63}\approx 0.349\label{exper}
\ee
As we can see the reconstructed matrix elements are far apart from the  
expected values.

Let's do the same for $D^T$. 
\be
D^T_{22}\,&=&\, \frac \pi{24} \int_{-\infty}^{\infty} d\k\,
 \k^2(\k^2+4) \frac{\cosh  \left( \frac {\pi\k}2 \right)}
{\left(\sinh\left( \frac {\pi\k}2 \right)\right)^2} \approx 2.665\label{D22}\\
 D^T_{24}\,&=&\, \frac \pi{480} \int_{-\infty}^{\infty} d\k\, 
(\k^2(\k^4-16))\frac{\cosh  \left( \frac {\pi\k}2\right)}
{\left(\sinh\left( \frac {\pi\k}2 \right)\right)^2} 
  \approx 0.5333\label{D42}\\
D^T_{33}\,&=&\, \frac \pi{96} \int_{-\infty}^{\infty} d\k\, 
(\k^4(\k^2+4))
 \frac{\cosh  \left( \frac {\pi\k}2\right)}
{\left(\sinh\left( \frac {\pi\k}2 \right)\right)^2} \approx 5.3333\label{D33}\\
D^T_{35}\,&=&\, \frac \pi{2880} \int_{-\infty}^{\infty} d\k\, 
(\k^4(\k^2-14)(\k^2+4))\frac{\cosh  \left( \frac {\pi\k}2 \right)}
{\left(\sinh\left( \frac {\pi\k}2 \right)\right)^2} 
 \approx 1.0667\label{D53}
\ee
We should have instead
\be
D_{22}=4,\quad\quad D_{42}=0,\quad\quad D_{33}=6,\quad\quad D_{53}=\frac 23\label{expD}
\ee
Also here we are far apart from the true values. 

We remark that if we do the same exercise for the $A$ and $C$ matrices of the matter sector
(see \cite{BMST}), we find a perfect coincidence between the original matrices and the 
ones reconstructed by means of the spectrum.

\subsection{The solution}

The idea is to apply the matrix $\tilde A$ not to the weight 2 basis, but 
to the weight -1 basis. I.e.
\be
\sum_{l=-1}^\infty \tilde V_l^{(-1)}(\k) \tilde A_{l\m} = {\mathfrak a}(\k)
\tilde V_{\m}^{(-1)}
\ee
This formula was proved in Appendix D3 of \cite{BMST}.
Then
\be
&&\int_{-\infty}^\infty d\k \tilde V^{(-1)}_{\m} (\k) 
{\mathfrak a}(\k)\tilde V_{\n}^{(2)}(\k) = \int_{-\infty}^\infty d\k 
\sum_{l=-1}^\infty V_l^{(-1)} (\k)\tilde A_{l\m} \tilde V_{\n}^{(2)}(\k)
\label{ricardo1}\\
&& = \sum_{a=-1,0,1} \tilde A_{a\m} \int_{-\infty}^\infty d\k \tilde 
V^{(-1)}_a (\k) \tilde V_n^{(2)}(\k)+ \sum_{l=2}^\infty \tilde A_{l\m}
\int_{-\infty}^\infty d\k \tilde V^{(-1)}_l (\k)\tilde V_{\n}^{(2)}(\k)\0
\ee
where barred indices denote `short' indices, i.e. $\m,\n\geq 2$. 
Now use the decomposition (see \cite{BeLove} and Appendix B of \cite{BMST})
\be
\tilde V^{(-1)}_a (\k) = \sum_{n=2}^\infty b_{a\n} \tilde V_{\n}^{(-1)}(\k)
\ee
One can easily obtain 
\be
b_{-1,2n+3}= (-1)^n (n+1),\quad\quad   b_{0,2n+2}= (-1)^n, \quad\quad
b_{1,2n+3}=(-1)^n (n+2)\label{b-1n}
\ee
Inserting these into (\ref{ricardo1}) we get
\be
\tilde A_{\n\m} = \int_{-\infty}^{\infty}  d\k \tilde V^{(-1)}_{\m} (\k)
{\mathfrak a}(\k) \tilde V_{\n}^{(2)}(\k) - \sum_{a=-1,0,1} b_{a\n} \tilde A_{a\m}
\label{Ricardo2}
\ee
Now the corrections to the
values obtained in (\ref{A22}-\ref{A35}) are easy to compute. For instance 
\be
\tilde A_{24} &=& -0.076 - \tilde A_{04} b_{02} 
=-0.076 + \frac 8{15}\approx 0.457\label{A22r}\\
\tilde A_{33} &=& 0.1524 - \tilde A_{-1,3}b_{-1,3}-\tilde A_{1,3}b_{1,3}=
  0.1524 -\frac 23 = - 0.5142\label{A33r}
\ee
and so on. 

As for $B$ the answer is easy since $B_{\n\m}=A_{\n\m}$. Notice that
the terms $B_{\n a}$ are different from $A_{a\n}$. These terms should also be 
considered as known terms.

We can reconstruct in a similar way also $D^T$.  For this we must apply 
$C$ to the -1 basis. This amounts to the same formulas above, with the  
substitution of $\tilde A$ with $C$ and ${\mathfrak a}(\k)$ with
${\mathfrak c}(\k)$.
Remember that $C_{\n\m}=D^T_{\n\m}$ for $\n,\m\geq 2$. In particular
\be
D^T_{\n\m}=C_{\n\m} = \int_{-\infty}^{\infty}  d\k \tilde V^{(-1)}_{\m} (\k)
{\mathfrak c}(\k) \tilde V_{\n}^{(2)}(\k) - \sum_{a=-1,0,1} b_{a\n} C_{a\m}
\label{Ricardo3}
\ee
For instance
\be
&&D_{22}^T =C_{22}= 2.6665- C_{02}\, b_{02} = 2.6665 + 2\,\frac 23 \approx
4\label{D22r}\\
&& D_{33}^T=C_{33}= 5.3333 -(C_{-1,3}\,b_{-1,3} +C_{1,3}\, b_{1,3})=
5.3333-\frac 23 + 2\frac 23\approx 6\label{D33r}\\
&&D_{35}^T=C_{35}= 1.0667 -(C_{-1,3}\,b_{-1,3}+ C_{1,3}\,b_{1,3})=
1.0667-\frac 25 \approx 0.6667=\frac 23\label{D35r}
\ee
and so on. $\tilde A_{a,\n}, \tilde B_{\n,a}$ and 
$C_{a,\n}$ and $C_{\n,a}$ will be referred to from now on as
{\it boundary terms}. Notice that $A_{a,n}= -C_{-a,n}$
and $C_{\n,-a}=B_{\n,a}$.

In the absence of a mathematical theorem we formulate the following: 

{\bf Heuristic rule}. {\it In order to reconstruct any matrix 
$A_{\n\m}=B_{\n\m}$ and $C_{\n\m}=D^T_{\n\m}$ from its eigenvalues, 
apply $A$ and $C$ to the weight -1 basis, separate the $\n,\m\geq 2$ part 
from the zero mode part and operate as in eq.(\ref{ricardo1}) above.
As for the matrix elements  $A_{a,n}$ and $B_{\n,a}$
they cannot be reconstructed from
the eigenvalues, but they have to be rather considered as known terms of 
the problem. We will refer to them as {\it boundary data}.}

Usually (for instance in the matter sector) we start from a matrix (for 
instance the matrices $\tilde A$ or $C$ of \cite{BMST}), diagonalize it and
determine the spectrum, i.e. eigenvalues and eigenvectors. Viceversa, 
starting from the latter, we can reconstruct the initial matrix using its
spectral representation.

In  the present case the situation is somewhat different. Given the 
matrices we can compute the spectrum (see section 5 of \cite{BMST}). 
Viceversa given the spectrum {\it and} the {\it boundary data} 
$A_{a,n}$ and $B_{\n,a}$ we can compute the matrices $\tilde A$ and $C$ and 
the related ones. This also means that, in order to determine the 
eigenvalue of a given diagonalizable matrix over the weight 2 basis, 
the $ss$ part of that matrix contains all the necessary information, 
but in order to reconstruct even its $ss$ part we have to know the 
action of that matrix over the weight -1 basis, i.e. we need the 
information stored in the latter.

It is clear that, with the above heuristic rule, it is possible to 
reconstruct, at least in principle, any matrix which can be expressed
as a series of products of $\tilde A,B,C,D^T$, in particular $\tilde\alpha(t)$. 
Unfortunately so far we have not been able
to produce a simple, manageable reconstruction algorithm.  
 
\section{Conclusion}

Let us return to the validity of (\ref{ghostwedge}) and reformulate the
question raised at the beginning of the previous section. 
In \cite{BMST} we wrote 
the LHS of this equation in the form (\ref{ghwed1}). We have shown above 
that the RHS of the latter has the form of a wedge state, and in fact we 
proved that once the squeezed state matrix $\tilde\alpha(\k,\frac{2-n}2)
\equiv \tilde\alpha_n$ 
there is diagonalized in the weight 2 basis, it coincides with the
(diagonalized) matrix that represents the n--th wedge state $|n\rangle$,
defined by the squeezed state (\ref{squeez}) whose matrix $T_n$ satisfies
the recursion relations (\ref{recur}). 

The next question to be answered is: in view of the discussion of the 
previous 
section, do also the matrices $\tilde\alpha_n$ coincide
with the matrices $T_n$ and, in particular, the matrix elements 
$(\tilde\alpha_n)_{a,m}$ 
with $(T_n)_{a,m}$ with $a=-1,0,1$ ?  
Remember that our reconstruction algorithm tells us that 
$(\tilde\alpha_n)_{a,\m}$ (beside the other matrix elements) is uniquely
determined by the spectrum of $\tilde \alpha_n$ {\it and} by the boundary data.
This is true in particular for $\tilde\alpha_3$, which was interpreted above
as $X$. We therefore expect that $(\tilde\alpha_3)_{n,m}$ 
coincides with
$X_{n,m}$. If this is so, solving (\ref{fund1},\ref{fund2}) for $X^\pm$, 
{\it we can, in principle, reconstruct the three 
strings vertex from the $A,B,C$ and $D^T$ matrices}. 
This vertex has precisely the features we have hypothesized in section 2 
and 3,
in particular the commutativity of the twisted matrices of Neumann 
coefficients (otherwise they would not be simultaneously diagonalized).

However the reconstruction of $X^\pm$ is not on the same footing as the 
reconstruction of $X$ (or $T$). For the latter, as we have seen, there 
exists a precise (though unwieldy) procedure to obtain it from the
$A,B,C$ and $D^T$ matrices. For $X^\pm$ instead we have to proceed on 
the basis of (\ref{fund1}) and (\ref{fund2}). To discuss this point let us 
introduce the following notation: for any matrix $M$ we represent by $M_{ee}$
the part of $M$ with both even entries, $M_{oo}$ the part with both odd 
entries, and accordingly $M_{eo}, M_{oe}$ with obvious meaning. 
From \cite{BMST} we know that all the matrices $A,B,C,D$ have vanishing
$_{eo}$ and $_{oe}$ parts. All matrices $T_n$, and in particular $X$ will 
therefore share the 
same property. We expect instead that $X^\pm_{eo}$ and $X^{\pm}_{oe}$ be 
nonvanishing. Remember that $X^+=\hat C X^- \hat C$. Therefore
\be
&&X^+_{ee} =X^-_{ee},\quad\quad X^+_{oo} =X^-_{oo}\label{eeoo}\\
&& X^+_{eo} =-X^-_{eo},\quad\quad X^+_{oe} =-X^-_{oe}\label{eooe}
\ee
Substituting these relations into (\ref{fund1}) we find
\be
&&X^+_{ee} =X^-_{ee}= \frac 12 (1_{ee} -X_{ee}) \label{ee}\\
&&X^+_{oo} =X^-_{oo}= \frac 12 (1_{oo} -X_{oo}) \label{oo}
\ee
Therefore both $X^\pm_{ee}$ and $X^\pm_{oo}$ are immediately derived
from $X$. From (\ref{fund2}) we get instead
\be
&&X^\pm_{eo}\,X^{\pm}_{oo}=X^{\pm}_{ee}\,X^{\pm}_{eo}\label{eooo}\\
&&X^\pm_{oe}\,X^{\pm}_{ee}=X^{\pm}_{oo}\,X^{\pm}_{oe}\label{oeee}
\ee
and
\be
X^\pm_{eo}\,X^\pm_{oe}= \frac 14 \left(1_{ee} +3X_{ee}\right)
\left(1_{ee} -X_{ee}\right)\label{quadratic}
\ee
and a parallel equation with $_o$ exchanged everywhere with $_e$.
This means that $X^\pm_{eo}$ and $X^\pm_{oe}$ are not determined 
algorithmically like $T_n$, but only by solving the quadratic equations
(\ref{quadratic}) subject to the commutativity relations
(\ref{eooo}, \ref{oeee}).

If the solution to such equations, as we expect, exists, this is a proof
of the validity of (\ref{ghostwedge}). In fact the analysis in section 3  
was carried out under the hypothesis that a vertex, with the properties 
illustrated in section 2.1 {\it and} with the twisted matrices of Neumann
coefficients commuting with $K_1$, existed. But we have just shown 
that such vertex can be deduced (reconstructed) directly from the LHS of
(\ref{ghostwedge}), in the sense we have just specified. 

The information we have extracted from the reconstruction path taken up 
in this paper is not conclusive. The existence proof of the three strings 
vertex, as we have just seen, is not complete. On the other hand 
the missing part in the proof is rather marginal and, what is more important,
our general characterization of the three strings vertex (section 2.1)
has not met any inconsistencies. This is reassuring in the prospect
of coping with the task of explicitly constructing the three strings 
vertex endowed with the above properties. 
 
\acknowledgments

We would like to thank Carlo Maccaferri for his comments and suggestions and
for his collaboration in the early stage of this work.
L.B. would like to thank the CBPF (Rio de Janeiro) and the YITP (Kyoto)
for their kind hospitality and support during this research.
This research was supported for L.B. by the Italian MIUR
under the program ``Superstringhe, Brane e Interazioni Fondamentali''.
D.D.T. was supported by the Science Research Center Program
of the Korea Science and Engineering Foundation through the Center for
Quantum Spacetime(CQUeST) of Sogang University with grant number
R11 - 2005 - 021. R.J.S.S. was supported by CNPq--Brasil.

\section*{Appendix}
\appendix

\section{The ghost Neumann coefficients}

In this Appendix we explicitly compute $\hat V_{nm}^{rs}$  
and $V_{nm}^{rs}$. We use the definitions (\ref{V3gh},\ref{V3gh'}).
The method is well--known: we express the propagator
$\ll c(z) b(w)\gg$  in two different ways, first as a 
CFT correlator and then in terms of $\hat V_3$ and we equate the two 
expressions after mapping them to the disk via the maps (\ref{fi})
However this recipe leaves several uncertainties.
We will fix them by requiring certain properties, in particular cyclicity,
consistency with the $bpz$ operation and commutativity of the twisted matrices
of Neumann coefficients (the reason for the latter will become clear later on).  

First we have to insert the three $c$ zero modes. One way 
is to insert them at different points $t_i$ and
use the correlator (\ref{leclaircorr})  
\be
\ll c(z) b(w)\gg_{(t_1,t_2,t_3)} &=& \langle 0|c(z)
b(w) c(t_1) c(t_2)c(t_3) |0\rangle \0\\
&=& \frac 1{z-w} \prod_{i=1}^3 \frac {t_i-z}{t_i-w}\,
(t_1-t_2)(t_1-t_3)(t_2-t_3)\label{leclaircorr}
\ee
So we have to compare
\be
\langle f_1 \circ c(t_1) \,f_2 \circ c(t_2)\,
f_3 \circ c(t_3) \, f_r \circ c^{(r)}(z) \, f_s \circ b^{(s)}(w) \rangle
\label{corrcb}
\ee
with 
\be
\langle \hat V_{3}|R(c^{(r)}(z)\, b^{(s)}(w)) |\omega\rangle_{123}
 \label{V3cb}
\ee
where $R$ denotes radial ordering. If :: denotes the natural normal ordering,
we have for instance
\be
R(c(z)\,b(w)) = \sum_{n,k} :c_n\, b_k: \, 
z^{-n+1} w^{-k-2} +\frac 1{z-w}\label{radial}
\ee
This should be inserted inside (\ref{V3cb}). Let us refer to 
the last term in (\ref{radial}) as the {\it ordering term}. 
We notice that the choice we have made for this term is rather 
arbitrary. What precisely has to be 
inserted in (\ref{V3cb}) depends   
also on the definition of the three strings vertex, therefore should be 
decided on the basis of a consistent definition of the latter. For the 
time being we continue on the basis of (\ref{radial}), later on we will 
introduce the necessary modifications.

To start with let us compute the ${\cal K}$ constant. We have
\be
&&\langle \hat V_{3}|\omega\rangle_{123}= {\cal K}
= \langle f_1 \circ c(t_1) \,f_2 \circ c(t_2)\,
f_3 \circ c(t_3) \rangle\nonumber\\
&& =\frac {(f_1(t_1) -f_2(t_2))\, (f_1(t_1) -f_3(t_3)) \,
(f_2(t_2) -f_3(t_3))} {f_1'(t_1)\,  f_2'(t_2)\,  f_3'(t_3)}
\label{calN}
\ee
Now
\be
&&\langle \hat V_{3}|R (c^{(r)}(z)\, b^{(s)}(w))
|\omega\rangle_{123}\0\\
&&= \langle \hat V_{3}|\sum_{n,k} :c^{(r)}_{n}\, b^{(s)}_{k}: \, z^{-n+1} w^{-k-2} 
+\frac 1{z-w}|\omega\rangle_{123} \0\\
&&
= {\cal K}\, \left( -\hat V^{sr}_{kn} \, z^{n+1} w^{k-2} + \frac {\delta^{rs}}
{z-w}\right)\label{cbcorr1}
\ee
On the other hand, from direct computation,
\be
&&\langle f_1 \circ c(t_1) \,f_2 \circ c(t_2)\,
f_3 \circ c(t_3) \, f_r \circ c^{(r)}(z) \, f_s \circ b^{(s)}(w) \rangle
\0\\
&&= \frac {(f_s'(w))^2}{f_r'(z)} \, \frac 1{f_r(z)-f_s(w)} \,
\frac {(f_1(t_1) -f_2(t_2))\, (f_1(t_1) -f_3(t_3)) \,
(f_2(t_2) -f_3(t_3))} {f_1'(t_1)\,  f_2'(t_2)\,  f_3'(t_3)}
\, \0\\
&&~~~~~~~~\cdot \prod_{i=1}^3\frac{ f_i(t_i) -f_r(z)}
{f_i(t_i) -f_s(w)}\label{cbcorr2}
\ee
Comparing the last two equations and using (\ref{calN}) we get
\be
\hat V_{kn}^{sr} &=& -\oint \frac {dz}{2\pi i}  \oint \frac {dw}{2\pi i}
\frac 1{z^{n+2}} \frac 1{w^{k-1}}\cdot \label{cbcorr3}\\
&&\cdot \left( 
\frac {(f_s'(w))^2}{f_r'(z)} \, \frac 1{f_r(z)-f_s(w)} \
\prod_{i=1}^3\frac {f_i(t_i) -f_r(z)}
 {f_i(t_i) -f_s(w)}-\frac {\delta^{rs}}
{z-w}\right)\0
\ee
After obvious changes of indices and variables we end up with
\be
\hat V_{nm}^{rs} &
=&\oint\frac{dz}{2\pi i}\oint\frac{dw}{2\pi i}\frac{1}{z^{n-1}}
\frac{1}{w^{m+2}}\label{cbcorr4}\\
&&\cdot \left(\frac{(f'_r(z))^2}{(f'_s(w))}\,
\frac{1}{f_r(z)-f_s(w)}\prod_{i=1}^3
\frac{f_s(w)-f_i(t_i)}{f_r(z)-f_i(t_i)}- \frac {\delta^{rs}}
{z-w}\right)\0
\ee
Now we make a definite choice for the insertions, that is 
we take $t_i\to \infty$. We remark that this choice leads to
simple formulas but remains anyhow arbitrary\footnote{We recall that zero mode 
insertions can be introduced also by means of the operator $Y(t)= \frac 12 
\partial^2 c(t) \partial c(t) c(t) $ instead of three different
$c(t_i)$. This has not lead so far to better results.}.

Since $f_i(\infty)= \alpha^{-i}$
we get
\be
\prod_{i=1}^3 \frac {f_i(t_i) -f_s(w)}
{f_i(t_i) -f_r(z)}= \frac {f(w)^3-1}{f(z)^3-1}\label{limitzi}
\ee

It is straightforward to check cyclicity
\be
\hat V_{nm}^{rs}=\hat V_{nm}^{r+1,s+1},\label{cyclgh'}
\ee
Moreover (by letting $z\rightarrow -z,\, w\rightarrow -w$)
\be
\hat V_{nm}^{rs}=(-1)^{n+m}\hat V_{nm}^{sr}\label{twistgh}
\ee
Now let us consider the decomposition
\be
\hat V_{nm}^{rs}=\frac{1}{3}(E_{nm}+\bar{\alpha}^{r-s} U_{nm}+
\alpha^{r-s}\bar{U}_{nm})\label{decompgh}
\ee
where
\be
E_{nm}&=&\oint\frac{dz}{2\pi i}
\oint\frac{dw}{2\pi i}{\N}_{nm}(z,w){\E}(z,w)\nonumber\\
U_{nm}&=&\oint\frac{dz}{2\pi i}
\oint\frac{dw}{2\pi i}{\N}_{nm}(z,w){\U}(z,w)\\
\bar{U}_{nm}&=&\oint\frac{dz}{2\pi i}
\oint\frac{dw}{2\pi i}{\N}_{nm}(z,w)\bar{\U}(z,w)\nonumber
\ee
and
\be
{\E}(z,w)\!&=&\!\frac{3f(z)f(w)}{f^3(z)-f^3(w)}\nonumber\\
{\U}(z,w)\!&=&\!\frac{3f^2(z)}{f^3(z)-f^3(w)}\nonumber\\
\bar{\U}(z,w)\!&=&\!\frac{3f^2(w)}{f^3(z)-f^3(w)}\0\\
{\N}_{nm}(z,w) \!&=&\! \frac 1{z^{n-1}}\frac 1{w^{m+2}}
(f'(z))^2 (f'(w))^{-1} \frac{f^3(w)-1}{f^3(z)-1}\0
\ee
After some elementary algebra, using
$f'(z)=\frac{4i}{3}\frac{1}{1+z^2}f(z)$, one finds
\be
E_{nm}&=&\oint\frac{dz}{2\pi i}\oint\frac{dw}{2\pi i}\frac{1}{z^{n+1}}
\frac{1}{w^{m+1}}
\Big{(}\frac{1}{1+zw}-\frac{w}{w-z}-\frac {z^2}w \frac 1{z-w}\Big{)} \0\\
&=& (-1)^n \delta_{nm} -\delta_{n,0}\delta_{m,0} 
-\delta_{n,1}\delta_{m,-1}\label{EUUbar}\\
U_{nm}&=&\oint\frac{dz}{2\pi i}\oint\frac{dw}{2\pi
i}\frac{1}{z^{n+1}}\frac{1}{w^{m+1}}\left[\frac{f(z)}{f(w)} 
\Big{(}\frac{1}{1+zw}
-\frac{w}{w-z}\Big{)}- \frac {z^2}w \frac 1{z-w} \right]\0\\
\bar{U}_{nm}&=&\oint\frac{dz}{2\pi i}\oint\frac{dw}{2\pi
i}\frac{1}{z^{n+1}}\frac{1}{w^{m+1}}\left[\frac{f(w)}{f(z)} 
\Big{(}\frac{1}{1+zw}
-\frac{w}{w-z}\Big{)}-\frac {z^2}w \frac 1{z-w} \right]\nonumber
\ee
In this way the ambiguities are eliminated. The $\delta_{n,m}$ in 
(\ref{EUUbar}) is for $n,m\geq 0$.

In a similar way one can compute the dual vertex, the right one.
One gets
\be
&&{}_{123}\langle \omega|R (I\circ c(z)\,
I\circ b(w))|V_3\rangle\0\\
&&=  {}_{123}\langle \omega|\sum_{n,k} 
(-1)^{k+n+1} 
:c^{(r)}_n b^{(s)}_k: \,z^{n+1}\,w^{k-2} + \frac {z^3} {w^3}
\frac {\delta^{rs}}{z-w}|V_3\rangle \0\\
&& ={\cal K} \left(\sum_{n,k} V_{kn}^{sr} (-1)^{n+m+1}  z^{n+1} w^{k-2}+
\frac {z^3} {w^3}\frac {\delta^{rs}}{z-w}\right)\label{cbcorr1'}
\ee

Equating now to (\ref{corrcb}) and 
repeating the same procedure as above we finally obtain
\be
(-1)^{n+m} V_{nm}^{rs}=\frac{1}{3}(E'_{nm}+\bar{\alpha}^{r-s} U'_{nm}+
\alpha^{r-s}\bar U'_{nm})\label{decompgh'}
\ee
where
\be
E'_{nm}&=&\oint\frac{dz}{2\pi i}\oint\frac{dw}{2\pi i}\frac{1}{z^{n+1}}
\frac{1}{w^{m+1}}
\Big{(}\frac{1}{1+zw}-\frac{w}{w-z}-\frac {w^2}z \frac 1{z-w}\Big{)}
\label{EUUbar'}\\
U'_{nm}&=&\oint\frac{dz}{2\pi i}\oint\frac{dw}{2\pi
i}\frac{1}{z^{n+1}}\frac{1}{w^{m+1}}\left[\frac{f(z)}{f(w)} 
\Big{(}\frac{1}{1+zw}
-\frac{w}{w-z}\Big{)}- \frac {w^2}z \frac 1{z-w} \right]\0\\
\bar U'_{nm}&=&\oint\frac{dz}{2\pi i}\oint\frac{dw}{2\pi
i}\frac{1}{z^{n+1}}\frac{1}{w^{m+1}}\left[\frac{f(w)}{f(z)} 
\Big{(}\frac{1}{1+zw}
-\frac{w}{w-z}\Big{)}-\frac {w^2}z \frac 1{z-w}\right]\nonumber
\ee

As we see, we have
\be
(-1)^{n+m} V_{nm}^{rs}= \hat V_{nm}^{rs}\label{bpz'}
\ee
except perhaps for the values of the labels both involving zero modes.
That the relation (\ref{bpz}) should hold for the full range
of the labels is instead a basic requirement. We will use also this, beside 
cyclicity and commutativity, in order to guess the final form of the vertex. 
 
Motivated by these requirements we introduce minor modifications
in the previous definitions. We start from the basic (\ref{EUUbar}) 
without the last term (the ordering term)
\be
E_{nm}&=&  \oint\frac{dz}{2\pi i}\oint\frac{dw}{2\pi i}\frac{1}{z^{n+1}}
\frac{1}{w^{m+1}}
\Big{(}\frac{1}{1+zw}-\frac{w}{w-z} \Big{)}\label{Enm}\\
U_{nm}&=&\oint\frac{dz}{2\pi i}\oint\frac{dw}{2\pi
i}\frac{1}{z^{n+1}}\frac{1}{w^{m+1}}\,\frac{f(z)}{f(w)} \Big{(}\frac{1}{1+zw}
-\frac{w}{w-z}\Big{)}\label{Unm}\\
\bar{U}_{nm}&=&\oint\frac{dz}{2\pi i}\oint\frac{dw}{2\pi
i}\frac{1}{z^{n+1}}\frac{1}{w^{m+1}}\,\frac{f(w)}{f(z)} \Big{(}\frac{1}{1+zw}
-\frac{w}{w-z}\Big{)}\label{Ubarnm}
\ee
Then we define the ordering term
\be
Z_{nm}= \oint\frac{dz}{2\pi i}\oint\frac{dw}{2\pi i}\frac{1}{z^{n+1}}
\frac{1}{w^{m+1}}\left(\frac{w}{w-z}- \frac 1{zw}\right)\label{Znm}
\ee
Next we define the matrices
\be
\EE= E+Z, \quad\quad \EU = U+Z,\quad\quad \bar \EU = \bar U +Z \label{EuEU}
\ee
which will be our basic ingredients.
The choice of $Z$ is made in such a way that $\EE= \hat C$. In fact 
\be
\EE_{nm}= \oint\frac{dz}{2\pi i}\oint\frac{dw}{2\pi i}\frac{1}{z^{n+1}}
\frac{1}{w^{m+1}}\left(\frac 1{1+zw}-\frac {1}{zw}\right)= (-1)^n\, 
\delta_{nm}\label{E=C}
\ee
for $n,m\geq -1$.

The double integrals in (\ref{Enm},\ref{Unm},\ref{Ubarnm}) are ambiguous 
in the range $-1\leq n,m\leq 1$, \cite{tope}. However, after the addition 
of the ordering term (\ref{Znm}) all ambiguities disappear.  

In conclusion we define the three strings ghost vertex as follows.
{\it With reference to (\ref{V3gh})
and  (\ref{V3gh'}) we set
\be
\hat V_{nm}^{rs}=\frac{1}{3}(\EE_{nm}+\bar{\alpha}^{r-s} \EU_{nm}+
\alpha^{r-s}\bar{\EU}_{nm})\label{hatVnmrs}
\ee
and
\be
V_{nm}^{rs}= (-1)^{n+m} \hat V_{nm}^{rs}, \quad\quad V^{rs} = \hat C 
\hat V^{rs} \hat C\label{Vnmrs}
\ee}

From the definition of $\EU$ and $\bar \EU$ it is easy to verify that
$\hat C \EU = \bar \EU \hat C$, where $\hat C$ denotes the twist matrix. 
We have seen above that $\EE \equiv \hat C$. 

Now using the method of (\cite{tope}) it is possible to show that
$\EU^2=1$ for $n,m\geq -1$. This implies that, beside
\be
X+X^++X^-=1\0
\ee
where $X=\hat CV^{rr}$, $X^+=\hat C V^{12},X^-= \hat CV^{21}$ , we have 
the commutativity property
\be
X^{rs}X^{r's'}= X^{r's'}X^{rs}\0
\ee
and
\be
X^+X^- = X^2 -X,\quad\quad X^2 +(X^+)^2+(X^-)^2=1\0
\ee
It should be stressed that all the $X^{rs}$ matrices are $(ll)$.

\section{Why we can use long square matrices}

Let us return to eqs.(\ref{T12},\ref{SigmaV}) and (\ref{norm12})
applied to wedge states, that is let us suppose $S_1=S_n$ and $S_2=S_m$.
Let us concentrate on eq.(\ref{T12}): 
the RHS can be understood in terms of a series expansion in which each 
monomial is the product of alternating lame matrices $X,X^\pm, T_n, T_m$, 
the rightmost and leftmost ones being $(ls)$. These matrices cannot be assumed 
to satisfy the identities of sec.2.1, in particular they cannot be assumed to 
commute. However let us apply any such monomial 
from the left to the above introduced weight 2 basis vector $V_s^{(2)}$:
\be
\ldots Y_{sl} Z_{ls} V_s^{(2)}\label{monomial}
\ee
Since the rightmost matrix $Z$ in the monomial
is $(ls)$, whatever matrix it is it is obvious that we can simply replace it 
with corresponding long square matrices and replace $V_s^{(2)}$ by
$V_l^{(2)}$. According to the discussion in section 3, the result of the 
application is the same extended vector multiplied by the matrix eigenvalue. 
This is obvious if the matrix in question is $X,T_n$ or $T_m$, as has been 
discussed above. If the rightmost matrix in the monomial is $X^\pm$ the 
same conclusion requires some comment. Since $X_{ls}^\pm$ can be
trivially replaced by a long square matrices applied to $V_l^{(2)}$, 
we are entitled to apply to $X_{ll}^\pm$ the identities of subsection 2.1. 
Therefore, using a well--known result, $X_{ll}^\pm$ can be expressed in terms 
of $X_{ll}$, and the result of the application of $X_{ll}^\pm$ to $V_l^{(2)}$
is $V_l^{(2)}$ multiplied by the matrix eigenvalue. 

The next to the rightmost matrix in the monomial we have picked up is 
of the type $Y_{sl}$. If $Y$ is $X,T_n$ or $T_m$ we can apply
to it the argument used in section 3 for $\alpha(t)$: we can replace them
with $Y_{ll}$ since, due to (\ref{eigenA=0}) and the consequences thereof,
the initial three elements of $Y_{ll} V_l^{(2)}$ are zero. The result
once again is the product of the eigenvalues of $Y_{ll}$ and of $Z_{ll}$
multiplying $V_l^{(2)}$. 

If, on the other hand, $Y_{sl}$ is $X_{sl}^\pm$, we can argue as follows.
The result of replacing $X_{sl}^\pm$ by $X_{ll}^\pm$ in front 
of $V_l^{(2)}$ is a vector with three more entries (corresponding as always 
to $n=-1,0,1$, if $n$ is the left label of $X^{\pm}$). However we can use
the same argument as above, remarking that $X_{ll}^\pm$ can be expressed 
in terms of $X_{ll}$. Therefore, due to (\ref{eigenA=0}) and its consequences,
we can conclude that these three additional entries are 0. Therefore
writing $X_{sl}^\pm V_l^{(2)}$ is tantamount to writing
$X_{ll}^\pm V_l^{(2)}$, and the result is once again $V_l^{(2)}$
multiplied by the product of the eigenvalues of $Y_{ll}$ and of $Z_{ll}$.

From this point on the argument is recursive and there is no need to
repeat it again. Re-summing the series we can conclude that in 
eq.(\ref{T12}) we can everywhere replace the matrices by the corresponding
long square ones. Analogous things can be repeated for eq. (\ref{norm12}).
This is our justification for enlarging the Fock
space at the beginning of this section. 

\section{Proof of eq.(3.5)}

We want to show that
\be
X=\sum_{n=2}^{\infty}{\tilde A}_{-1,n}V^{(2)}_n(\k)=0\label{X=0}
\ee
Set $n=2l+1$. Then
\be
{\tilde A}_{-1,2l+1}={2(-1)^{l}\over 2l+1}\0
\ee
are the only non--vanishing matrix elements. Define
\be
F(z)=\sum_{l=1}^{\infty}{2(-1)^{l}\over 2l+1}V^{(2)}_{2l+1}z^{2l+1}\label{F}
\ee
so that $X=F(1)$ and $F(0)=0$. We get
\be
{dF\over dz}&=&\sum_{l=1}^{\infty}{2(-1)^{l}}V^{(2)}_{2l+1}z^{2l}=
iz\left(f^{(2)}_k(iz)-f^{(2)}_k(-iz)\right)\0\\
&=&{iz\over (1-z^2)^2}\left(\left({1+z\over 1-z}\right)^\xi-
\left({1+z\over 1-z}\right)^{-\xi}\right),~~~~~~\xi={i\k\over 2}\0
\ee
Therefore
\be
F(1)&=&\int_{0}^{1}dz (iz)\left((1+z)^{\xi-2}(1-z)^{-\xi-2}-
(1+z)^{-\xi-2}(1-z)^{\xi-2}\right)\0\\
&=&{i\over\xi(1+\xi)}F(2-\xi,2,1-\xi,-1)+
{i\over\xi(1-\xi)}F(2+\xi,2,1+\xi,-1)\0
\ee
Using eq.(C.2) of I one gets
\be
F(2-\xi,2,1-\xi,-1)=\frac 14 \frac \xi{\xi-1} \0
\ee
Therefore one can easily show that
\be
{i\over\xi(1+\xi)}F(2-\xi,2,1-\xi,-1)=-{i\over 4(1-\xi^2)}\0\\
{i\over\xi(1-\xi)}F(2+\xi,2,1+\xi,-1)={i\over 4(1-\xi^2)}\0
\ee
and $F(1)=0$.

Next we want to show
\be
Y=\sum_{n=2}^{\infty}{\tilde A}_{0,n}V^{(2)}_n(k)=0\label{Y=0}
\ee
This time put $n=2l$
\be
{\tilde A}_{0,2l}=(-1)^{l+1}\left({1\over 2l+1}+{1\over 2l-1}\right)\0
\ee
Define
\be
F(z)&=&\sum_{l=1}^{\infty}{(-1)^{l+1}\over 2l+1}V^{(2)}_{2l}z^{2l+1}\label{F1}\\
G(z)&=&\sum_{l=1}^{\infty}{(-1)^{l+1}\over 2l-1}V^{(2)}_{2l}z^{2l-1}\label{G1}
\ee
so that $F(1)+G(1)=Y$ and $F(0)+G(0)=0$. We get 
\be
{dF\over dz}&=&\sum_{l=1}^{\infty}{(-1)^{l+1}}V^{(2)}_{2l}z^{2l}=
{z^2\over 2}\left(f^{(2)}_k(iz)+f^{(2)}_k(-iz)\right)\0\\
&=&{z^2\over 2(1-z^2)^2}\left(\left({1+z\over 1-z}\right)^\xi+
\left({1+z\over 1-z}\right)^{-\xi}\right)\0
\ee
and
\be
{dG\over dz}&=&\sum_{l=1}^{\infty}{(-1)^{l+1}}V^{(2)}_{2l}z^{2l-2}=
{1\over 2}\left(f^{(2)}_k(iz)+f^{(2)}_k(-iz)\right)\0\\
&=&{1\over 2(1-z^2)^2}\left(\left({1+z\over 1-z}\right)^\xi+
\left({1+z\over 1-z}\right)^{-\xi}\right),\0
\ee
which give
\be
F(1)&=&\int_{0}^{1}dz {z^2\over 2}\left((1+z)^{\xi-2}(1-z)^{-\xi-2}+
(1+z)^{-\xi-2}(1-z)^{\xi-2}\right)\0\\
&=&{1\over\xi(1+\xi)(1-\xi)}F(2-\xi,3,2-\xi,-1)-
{1\over\xi(1-\xi)(1+\xi)}F(2+\xi,3,2+\xi,-1)=0\label{F(1)}
\ee
and
\be
G(1)&=&\int_{0}^{1}dz {1\over 2}\left((1+z)^{\xi-2}(1-z)^{-\xi-2}+
(1+z)^{-\xi-2}(1-z)^{\xi-2}\right)\0\\
&=&-{1\over(1+\xi)}F(2-\xi,1,-\xi,-1)-{1\over(1-\xi)}F(2+\xi,1,\xi,-1)=0
\label{G(1)}
\ee
These identities can be obtained by means of well--known relations valid for
the hypergeometric functions, such as those in Appendix C of \cite{BMST}.

Lastly we want to show that
\be
Z=\sum_{n=2}^{\infty}{\tilde A}_{1,n}V^{(2)}_n(k)=0\label{Z=0}
\ee
Setting $n=2l+1$ one realizes that
\be
{\tilde A}_{1,2l+1}={2(-1)^{l+1}\over 2l+1}=-{\tilde A}_{-1,2l+1}
\ee
So $Z=-X=0$.


\begin{thebibliography}{99}


\bibitem{BMST} L.~Bonora, C.~Maccaferri, R.~J.~Scherer Santos and D.~D.~Tolla,
{\it Ghost story. I. Wedge states in the oscillator formalism,} arXiv:0706.1025 [hep-th].

\bibitem{Schnabl05}
  M.~Schnabl,
  {\it Analytic solution for tachyon condensation in open string field theory,}
  Adv.\ Theor.\ Math.\ Phys.\  {\bf 10} (2006) 433
  [arXiv:hep-th/0511286].
 
\bibitem{Okawa1}
  Y.~Okawa,
 {\it Comments on Schnabl's analytic solution for tachyon condensation in
  Witten's open string field theory,}
  JHEP {\bf 0604} (2006) 055
  [arXiv:hep-th/0603159].

\bibitem{Ellwood:2006ba}
  I.~Ellwood and M.~Schnabl,
 {\it Proof of vanishing cohomology at the tachyon vacuum,}
  JHEP {\bf 0702} (2007) 096
  [arXiv:hep-th/0606142].

\bibitem{RZ06}
  L.~Rastelli and B.~Zwiebach,
  {\it Solving open string field theory with special projectors,}
arXiv:hep-th/0606131.

\bibitem{ORZ}
  Y.~Okawa, L.~Rastelli and B.~Zwiebach,
 {\it Analytic solutions for tachyon condensation with general projectors,}
  arXiv:hep-th/0611110.

\bibitem{Schnabl:2007az}
  M.~Schnabl,
 {\it Comments on marginal deformations in open string field theory,}
  arXiv:hep-th/0701248.

\bibitem{KORZ}
  M.~Kiermaier, Y.~Okawa, L.~Rastelli and B.~Zwiebach,
 {\it Analytic solutions for marginal deformations in open string field 
theory,}
  arXiv:hep-th/0701249.

\bibitem{Fuchs1}
E.~Fuchs and M.~Kroyter,
{\it Schnabl's L(0) operator in the continuous basis,}
JHEP {\bf 0610} (2006) 067
[arXiv:hep-th/0605254].

\bibitem{Fuchs2}
  E.~Fuchs and M.~Kroyter,
 {\it Universal regularization for string field theory,}
  JHEP {\bf 0702} (2007) 038
  [arXiv:hep-th/0610298].

\bibitem{Fuchs3}
  E.~Fuchs, M.~Kroyter and R.~Potting,
  {\it Marginal deformations in string field theory,}
   arXiv:0704.2222 [hep-th].

\bibitem{Fuchs0}
  E.~Fuchs and M.~Kroyter,
 {\it On the validity of the solution of string field theory,}
  JHEP {\bf 0605} (2006) 006
  [arXiv:hep-th/0603195].

\bibitem{Okawa2}
  Y.~Okawa,
{\it Analytic solutions for marginal deformations in open superstring field
  theory,}
  arXiv:0704.0936 [hep-th].

\bibitem{Okawa3}
  Y.~Okawa,
{\it Real analytic solutions for marginal deformations in open superstring field
  theory,}
  arXiv:0704.3612 [hep-th].

\bibitem{Samu}
S.~Samuel,
{\it The Ghost Vertex In E. Witten's String Field Theory},
Phys.\ Lett.\ B {\bf 181} (1986) 255.

\bibitem{CST}
E.Cremmer,A.Schwimmer, C.Thorn, {\it "The vertex function in Witten's
formulation of string field theory}, Phys.Lett. {\bf 179B} (1986) 57.

\bibitem{GJ1} D.J.Gross and A.Jevicki, {\it Operator Formulation
of Interacting String Field Theory}, Nucl.Phys. {\bf B283} (1987) 1.

\bibitem{GJ2} D.J.Gross and A.Jevicki, {\it Operator Formulation
of Interacting String Field Theory, 2}, Nucl.Phys. {\bf B287} (1987) 225.

\bibitem{Ohta}
N.~Ohta,
{\it ``Covariant Interacting String Field Theory In The Fock Space Representation,''}
Phys.\ Rev.\ D {\bf 34} (1986) 3785
[Erratum-ibid.\ D {\bf 35} (1987) 2627].
\bibitem{leclair} A.Leclair, M.E.Peskin, C.R.Preitschopf, {\it String Field
Theory on the Conformal Plane. (I) Kinematical Principles},
Nucl.Phys. {\bf B317} (1989) 411.

\bibitem{RSZ2} L.Rastelli, A.Sen and B.Zwiebach,
{\it Classical solutions in string field theory around the tachyon vacuum,}
  Adv.\ Theor.\ Math.\ Phys.\  {\bf 5} (2002) 393
  [arXiv:hep-th/0102112].
  
\bibitem{RSZ3} L.Rastelli, A.Sen and B.Zwiebach, {\it Half-strings,
Projectors, and Multiple D-branes in Vacuum String Field Theory},
JHEP {\bf 0111} (2001) 035 [hep-th/{0105058}].
 
\bibitem{tope}
L.~Bonora, C.~Maccaferri, D.~Mamone and M.~Salizzoni,
{\it ``Topics in string field theory,''}
arXiv:hep-th/0304270.
 
\bibitem{RSZ1} L.Rastelli, A.Sen and B.Zwiebach, {\it Star Algebra
Spectroscopy}, [\hepth{0111281}].

\bibitem{GRSZ2}
  D.~Gaiotto, L.~Rastelli, A.~Sen and B.~Zwiebach,
 {\it Star algebra projectors,}
  JHEP {\bf 0204} (2002) 060
  [arXiv:hep-th/0202151].

\bibitem{GRSZ1}
  D.~Gaiotto, L.~Rastelli, A.~Sen and B.~Zwiebach,
  {\it Ghost structure and closed strings in vacuum string field theory,}
  Adv.\ Theor.\ Math.\ Phys.\  {\bf 6} (2003) 403
  [arXiv:hep-th/0111129].

\bibitem{Schnabl2}   M.~Schnabl,
  {\it Wedge states in string field theory,}
  JHEP {\bf 0301} (2003) 004
  [arXiv:hep-th/0201095].

\bibitem{MM} C.~Maccaferri and D.~Mamone,
 {\it Star democracy in open string field theory,}
  JHEP {\bf 0309} (2003) 049
  [arXiv:hep-th/0306252].

\bibitem{Belov1}
  D.~M.~Belov,
  {\it Witten's ghost vertex made simple (bc and bosonized ghosts),}
  Phys.\ Rev.\  D {\bf 69} (2004) 126001
  [arXiv:hep-th/0308147].

\bibitem{Belov2}
  D.~M.~Belov and C.~Lovelace,
  {\it Star products made easy,}
  Phys.\ Rev.\  D {\bf 68} (2003) 066003
  [arXiv:hep-th/0304158].

\bibitem{BeLove}
  D.~M.~Belov and C.~Lovelace, {\it Unpublished}

\bibitem{Oku2} K.Okuyama, {\it Ghost Kinetic Operator of Vacuum
String Field Theory}, JHEP {\bf 0201} (2002) 027 [hep-th/{0201015}].

\bibitem{HKw} H.Hata and T.Kawano, {\it Open string states around
a classical solution in vacuum string field theory},
JHEP {\bf 0111} (2001) 038 [hep-th/{0108150}].

\bibitem{KP} V.A.Kostelecky and R.Potting, {\it Analytical construction
of a nonperturbative vacuum for the open bosonic string},
Phys.\ Rev.\ D {\bf 63} (2001) 046007
[hep-th/{0008252}].

\bibitem{FKM} E.Fuchs, M.Kroyter and A.Marcus, {\it Squeezed States 
Projectors in String Field Theory}, JHEP {\bf 0209} (2002) 022 [hep-th/{0207001}].


\bibitem{Furu}
  K.~Furuuchi and K.~Okuyama,
  {\it ``Comma vertex and string field algebra,''}
  JHEP {\bf 0109} (2001) 035
  [arXiv:hep-th/0107101].
 
\bibitem{Kishimoto}
  I.~Kishimoto,
  {\it Some properties of string field algebra,}
  JHEP {\bf 0112} (2001) 007
  [arXiv:hep-th/0110124].



\end{thebibliography}
\end{document}